\documentclass{aip-cp}

\usepackage[numbers]{natbib}
\usepackage{rotating}
\usepackage{epsf}
\usepackage{amssymb}
\usepackage{amsfonts}
\usepackage{psfrag,epsfig,graphicx}

\begin{document}

\title{Impact Factor for Exclusive Diffractive Dijet Production with NLO Accuracy}

\author[aff1]{R.~Boussarie\corref{cor1}}
\author[aff2]{A.~V.~Grabovsky\corref{cor2}}
\author[aff3]{L.~Szymanowski\corref{cor3}}
\author[aff4]{S.~Wallon\corref{cor4}}

\affil[aff1]{Institute of Nuclear Physics, Polish Academy of Sciences, Radzikowskiego 152, PL-31-342 Krak\'ow, Poland}
\affil[aff2]{Novosibirsk State University,
2 Pirogova street, Novosibirsk, Russia\\
Theory division, Budker Institute of Nuclear Physics,
11 Lavrenteva avenue, Novosibirsk, Russia}
\affil[aff3]{National Centre for Nuclear Research (NCBJ), Ho\.za 69, 00-681 Warsaw, Poland}
\affil[aff4]{Laboratoire de Physique Th\'{e}orique (UMR 8627), CNRS, Univ. Paris-Sud, \\
Universit\'{e} Paris-Saclay, 91405 Orsay Cedex, France \\ UPMC, Universit\'{e} Paris 06, Facult\'{e} de Physique, 4 place Jussieu, 75252 Paris, France \\}
\corresp[cor1]{renaud.boussarie@ifj.edu.pl}
\corresp[cor2]{a.v.grabovsky@inp.nsk.su}
\corresp[cor3]{lech.szymanowski@ncbj.gov.pl}
\corresp[cor4]{samuel.wallon@th.u-psud.fr}

\maketitle

\begin{abstract}
Relying on the shockwave approach, we present the main steps of the computation of the impact factor for the exclusive diffractive photo- or electro- production of a forward dijet with NLO accuracy. We provide details of the cancellation mechanisms for all the divergences which appear in the intermediate results.
\end{abstract}

\section{INTRODUCTION}

We report on our result on the one loop $\gamma^{(*)}\to q\bar{q}$ impact factor, a first step toward a complete next-to-leading-order (NLO) description of many inclusive or exclusive diffractive processes, either in the linear BFKL ~\cite{Fadin:1975cb, Kuraev:1976ge, Kuraev:1977fs, Balitsky:1978ic} or the non-linear color glass condensate (CGC) approaches~\cite{JalilianMarian:1997jx,JalilianMarian:1997gr,JalilianMarian:1997dw,JalilianMarian:1998cb,Kovner:2000pt,Weigert:2000gi,Iancu:2000hn,Iancu:2001ad,Ferreiro:2001qy}. Diffraction can be theoretically described either
using
a {\em resolved} Pomeron contribution, or using a {\em direct} Pomeron contribution involving the coupling of a Pomeron with the diffractive state. Our results are based on the latter. We show in particular that our result for the real contributions to the leading order (LO) $\gamma^{(*)}\to q\bar{q} g$ impact factor and to the next-to-leading order (NLO) 
$\gamma^{(*)}\to q\bar{q}$ impact factor allows one to extract the finite part of the NLO impact factor for diffractive dijet production.

\section{INTRODUCTION TO THE SHOCKWAVE FORMALISM}

Our calculation relies on Balitsky's QCD shockwave formalism~\cite{Balitsky:1995ub, Balitsky:1998kc, Balitsky:1998ya, Balitsky:2001re}.
We introduce two lightcone vectors
$n_{1}$ and $n_{2}$%
\begin{equation}
\label{Sudakov-basis}
n_{1} \equiv \left(  1,0,0,1\right)  ,\quad n_{2} \equiv \frac{1}{2}\left(  1,0,0,-1\right)
,\quad n_{1}^{+}=n_{2}^{-}=n_{1} \cdot n_{2}=1 \,,
\end{equation}
and the Wilson lines as 
\begin{equation}
U_{i}^\eta=U_{\vec{z}_{i}}^\eta = T \exp\left[{ig\int_{-\infty
}^{+\infty}b_{\eta}^{-}(z_{i}^{+},\vec{z}_{i}) \, dz_{i}^{+}}\right]\,.
\label{WL}%
\end{equation}
The operator $b_{\eta}^{-}$ is the external shockwave field built from slow gluons 
whose momenta are limited by the longitudinal cutoff $e^\eta p_\gamma^+$, where $\eta$ is an arbitrary negative parameter:
\begin{equation}
b_{\eta}^{-}=\int\frac{d^{4}p}{\left(  2\pi\right)  ^{4}}e^{-ip \cdot z}b^{-}\left(
p\right)  \theta\left(e^\eta-\frac{|p^{+}|}{p^+_\gamma}\right)\,,\label{cutoff}%
\end{equation}
where $p_\gamma$ is the momentum of the photon, which has a large component in the $+$ direction.
We use the lightcone gauge
$\mathcal{A}\cdot n_{2}=0,$
with $\mathcal{A}$ being the sum of the external field $b_\eta$ and the quantum field
$A_\eta$%
\begin{equation}
\mathcal{A}^{\mu} = A^{\mu}_\eta+b_\eta^{\mu},\;\;\;\;\;\;\;\;\;\quad b_\eta^{\mu}\left(  z\right)  =b_\eta^{-}(z^{+},\vec{z}\,) \,n_{2}%
^{\mu}=\delta(z^{+})B_\eta\left(  \vec{z}\,\right)  n_{2}^{\mu}\,,\label{b}%
\end{equation}
where
$B_\eta(\vec{z})$ is a profile function and the form for $b_\eta$ is valid in the small $x$ limit considered here.
From the Wilson lines, we define the dipole operator and its Fourier transforms as follows:
\begin{eqnarray}
\mathbf{U}_{ij}^\eta  \equiv  1 - \frac{1}{N_c} \mathrm{Tr}(U_i^\eta U_j^{\eta\dagger}), \ \
\tilde{\mathbf{U}}_{ij}^\eta  \equiv  \! \! \int \!\! d^d\vec{z}_i \, d^d\vec{z}_j \, e^{-i(\vec{p}_i \cdot \vec{z}_i) -i(\vec{p}_j \cdot \vec{z}_j)} \mathbf{U}_{ij}^\eta, \ \
\widetilde{\mathbf{U}_{ik}^\eta\mathbf{U}_{kj}^\eta} \equiv  \! \! \int \! \! d^d\vec{z}_i \, d^d\vec{z}_j \, d^d\vec{z}_k e^{-i(\vec{p}_i \cdot \vec{z}_i) -i(\vec{p}_j \cdot \vec{z}_j) -i(\vec{p}_k \cdot \vec{z}_k)} \mathbf{U}^\eta_{ik}\mathbf{U}^\eta_{kj}\,.\!\!\!
\end{eqnarray}
To get a physical amplitude, one should act with these operators on the incoming and outgoing states of the target. For a diffractive $\gamma^{(\ast)}(p_\gamma) P(p_0) \rightarrow X(p_X) P^\prime(p_0^\prime)$ process, the following matrix elements will be involved:
\begin{equation}
\mathbf{W}^\eta \rightarrow \langle P^\prime (p_0^\prime) \vert T(\mathbf{W}^\eta) \vert P(p_0) \rangle,
\end{equation}
where $\mathbf{W}^\eta$ is an operator built from the Wilson lines. In our case, there are two possibilities for $\mathbf{W}^\eta$: either a dipole operator $\mathbf{W}^\eta = \mathbf{U}_{ij}^\eta$, or a double-dipole operator\footnote{In the t'Hooft limit $N_c^{-2} \rightarrow 0$ or in the mean field approximation, the matrix elements for the double dipole operators can be written as the product of the matrix elements for two dipole operators.} $\mathbf{W}^\eta = \mathbf{U}_{ik}^\eta \mathbf{U}_{kj}^\eta$. We will write $\mathbf{W}$ rather than $\mathbf{W}^\eta$ for readability.

\section{IMPACT FACTOR FOR THE $\gamma^{(\ast)}\rightarrow q\bar{q}$ TRANSITION}

At leading order, the evaluation of the diagram contributing to the impact factor for the $\gamma^{\ast}\rightarrow q\bar{q}$ transition is straightforward.
After the projection on the color singlet state and the subtraction of the contribution without interaction with the external field, the contribution of this diagram can be written in the momentum space as the following convolution of Wilson line operators with the impact factor:
\begin{equation}
M_{LO}^{q\bar{q}} = \varepsilon_\alpha \int d\vec{p}_{1}d\vec{p}_{2} \, \delta(\vec{p}_{q1} + \vec{p}_{\bar{q}2}) \, \delta(p_q^+ + p_{\bar{q}}^+ -p_\gamma^+) \, \Phi_0^\alpha (\vec{p}_1,\,\vec{p}_2)\tilde{\mathbf{U}}_{12}\,,
\label{LOconv}%
\end{equation}
where we denoted $p_{ij} \equiv p_i-p_j$, and where $p_q$ (resp. $p_{\bar{q}}$) is the momentum of the outgoing quark (resp. antiquark).

\begin{figure}
\hspace{-0cm}\rotatebox{-90}{\psfrag{D}{1}
\includegraphics[scale=0.55]{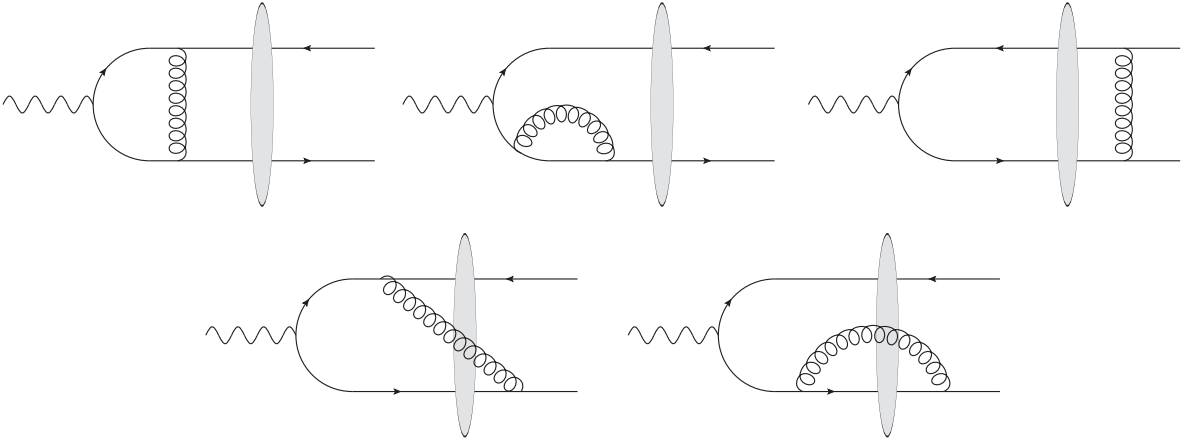}}
\caption{Example of virtual corrections for dijet production. The radiated gluon either crosses the shockwave (second line) or not (first line).}
\label{fig:virtual}
\end{figure}

The virtual corrections to the $\gamma^{(*)} \to q \bar{q}$ transition involve two kinds of contributions.
Examples of diagrams contributing to these virtual corrections 
are shown in Figure~\ref{fig:virtual}. The convolution is similar to the leading order result, but it involves more Wilson line operators:
\begin{eqnarray}
& & M_{NLO}^{q\bar{q}} = \varepsilon_\alpha \int \! d^d\vec{p}_1 \, d^d\vec{p}_2 \, d^d\vec{p}_3 \, \delta(\vec{p}_{q1} + \vec{p}_{\bar{q}2}-\vec{p}_3) \delta(p_q^++p_{\bar{q}}^+-p_\gamma^+) \label{NLOconv} \\
 & \times & \left\{ \left( \frac{N_c^2-1}{N_c} \right) \tilde{\mathbf{U}}_{12} \, \delta(\vec{p}_3)\left[ \Phi_{V_1}^\alpha + \Phi_{V_2}^\alpha \right] \nonumber \right. + \, N_c \left. \left( \widetilde{\mathbf{U}_{13}\mathbf{U}_{32}} + \tilde{\mathbf{U}}_{13} + \tilde{\mathbf{U}}_{32} - \tilde{\mathbf{U}}_{12} \right) \Phi_{V_2}^\alpha\right\}, \nonumber
\end{eqnarray} 
where $\Phi_{V_1}^\alpha = \Phi_{V_1}^\alpha(\vec{p}_1,\,\vec{p}_2)$ is obtained\footnote{Part of the contributions in $\Phi_{V_1}$ were also obtained in \cite{Beuf:2016wdz}.} from  diagrams of the type of the ones on the first line of Figure~\ref{fig:virtual} and $\Phi_{V_2}^\alpha = \Phi_{V_2}^\alpha(\vec{p}_1,\,\vec{p}_2,\,\vec{p}_3)$ is obtained from  diagrams of the type of the ones on the second line of Figure~\ref{fig:virtual}.

Several divergences appear in each of the terms in Equation~(\ref{NLOconv}): $\Phi_{V_1}^\alpha$ contains soft, collinear, soft and collinear, and UV divergences, while $\Phi_{V_2}^\alpha$ contains a rapidity divergence. In the shockwave formalism and in lightcone gauge, one has to combine the cutoff prescription $p^+ < e^\eta p_\gamma^+$
and the transverse dimensional regularization $d=2+2\epsilon$.   
The rapidity divergence in $\Phi_{V_2}$ is canceled via the use of the B-JIMWLK evolution equation for the dipole operator: evolving the dipole operator in the leading order convolution~(\ref{LOconv}) w.r.t. the longitudinal cutoff from the arbitrary $e^\eta p_\gamma^+$ to a more physical divide $e^{\eta_0} p_\gamma^+$, which will serve as a factorization scale which separates the upper and lower impact factors, allows one to cancel the dependence on $\eta$ in $\Phi_{V_2}$ and get a finite expression for the double-dipole contribution to the NLO impact factor. In momentum space and in $d+2$ dimensions, the evolution equation is given by:
\begin{eqnarray}
& & \hspace*{-.65cm} \frac{\partial{\mathbf{\tilde{U}}_{12}^{\eta}}}{\partial\mathrm{log}\eta} = 2\alpha_{s}N_{c}\mu^{2-d}\int\frac{d^{d}\vec{k}_{1}d^{d}\vec{k}_{2}d^{d}\vec{k}_{3}}{\left(2\pi\right)^{2d}}\delta\left(\vec{k}_{1}+\vec{k}_{2}+\vec{k}_{3}-\vec{p}_{1}-\vec{p}_{2}\right)\left(\widetilde{\mathbf{U}_{13}^\eta\mathbf{U}}_{32}^\eta+\tilde{\mathbf{U}}_{13}^\eta+\tilde{\mathbf{U}}_{32}^\eta-\tilde{\mathbf{U}}_{12}^\eta\right) \nonumber \\
\hspace*{-.75cm} & \times & \hspace*{-.25cm}  \left[2\frac{(\vec{k}_{1}-\vec{p}_{1}).(\vec{k}_{2}-\vec{p}_{2})}{(\vec{k}_{1}-\vec{p}_{1})^{2}(\vec{k}_{2}-\vec{p}_{2})^{2}}+\frac{\pi^{\frac{d}{2}}\Gamma\left(1-\frac{d}{2}\right)\Gamma^{2}\left(\frac{d}{2}\right)}{\Gamma\left(d-1\right)}\left(\frac{\delta(\vec{k}_{2}-\vec{p}_{2})}{\left[(\vec{k}_{1}-\vec{p}_{1})^{2}\right]^{1-\frac{d}{2}}}+\frac{\delta(\vec{k}_{1}-\vec{p}_{1})}{\left[(\vec{k}_{2}-\vec{p}_{2})^{2}\right]^{1-\frac{d}{2}}}\right)\right].\, \label{JIMWLK}
\end{eqnarray}
The divergences in $\Phi_{V_1}$ must be canceled by combining such terms with the associated real corrections to form a physical cross section. The first step to compute such a cross section is to use a jet algorithm in order to cancel the soft and collinear divergence. By using the jet cone algorithm in the small cone limit, as used in~\cite{Ivanov:2004pp}, we proved that such a cancellation occurs. 

The remaining divergence can be expressed by factorizing the leading order cross section:
\begin{equation}
d\sigma_{Vdiv}^{jets} = (N_V+N_V^\ast) \, d\sigma_{LO}^{jets}, \label{VirtualDiv}
\end{equation}
where $N_V$ is extracted from the divergent part of the virtual amplitude.
This contribution must be combined with real corrections from the $\gamma^{(\ast)} \rightarrow q\bar{q}g$ impact factor.

\section{IMPACT FACTOR FOR THE $\gamma^{(\ast)}\rightarrow q\bar{q}g$ TRANSITION}

The convolution for the $\gamma^{(\ast)}\rightarrow q\bar{q}g$ impact factor is very similar to the one for the NLO $\gamma^{(\ast)}\rightarrow q\bar{q}$ impact factor:
\begin{eqnarray}
M^{q\bar{q}g} & = & \varepsilon_\alpha \int \! d^d\vec{p}_1 \, d^d\vec{p}_2 \, d^d\vec{p}_3 \, \delta(\vec{p}_{q1} + \vec{p}_{\bar{q}2}+\vec{p}_{g3}) \delta(p_q^++p_{\bar{q}}^++p_g^+-p_\gamma^+) \label{NLOconvbis} \\
 & \times & \left\{ \left( \frac{N_c^2-1}{N_c} \right) \left[ \Phi_{R_1}^\alpha + \Phi_{R_2}^\alpha \right] \tilde{\mathbf{U}}_{12} \, \delta(\vec{p}_3)  + \, N_c  \left( \widetilde{\mathbf{U}_{13}\mathbf{U}_{32}} + \tilde{\mathbf{U}}_{13} + \tilde{\mathbf{U}}_{32} - \tilde{\mathbf{U}}_{12} \right) \Phi_{R_2}^\alpha \right\}, \nonumber
\end{eqnarray} 
where $\Phi_{R_1}=\Phi_{R_1}(\vec{p}_1, \, \vec{p}_2)$ and $\Phi_{R_2}=\Phi_{R_2}(\vec{p}_1, \, \vec{p}_2,\,\vec{p}_3)$ are obtained by computing respectively the two diagrams in which the emitted gluon does not cross (resp. crosses) the shockwave, as described in \cite{Beuf:2011xd,Boussarie:2014lxa} and \cite{Ayala:2016lhd}. 

When considering our exclusive cross section, the real contributions are those where the additional gluon is either collinear to the quark or to the antiquark, so that they form a single jet, or too soft to be detected \textit{i.e.} with an energy which is lower than a typical energy resolution $E$. The contribution from the soft gluon to the dijet cross section can be written with a very simple form:
\begin{equation}
d\sigma_{soft}^{q\bar{q}g} =  \alpha_s \left(\frac{N_c^2-1}{2N_c}\right)\int \frac{dp_g^+}{p_g^+}\frac{d^d\vec{p}_g}{(2\pi)^d} \left\vert \frac{p_q}{(p_q \cdot p_g)} - \frac{p_{\bar{q}}}{(p_{\bar{q}} \cdot p_g)} \right\vert ^2 d\sigma_{LO}^{jets} , \label{RealSoft}
\end{equation}
where the integration is performed in the $p_g$-phase space region where $p_g^+ + \frac{\vec{p}_g^2}{p_g^+} < 2E$. 

The collinear contribution also has a simple form, in terms of the jet variables. For example when the gluon is collinear to the quark one gets: 
\begin{equation}
d\sigma^{(qg),\bar{q}} = \alpha_s \left(\frac{N_c^2-1}{2N_c}\right) N_J \, d\sigma_{LO}^{jets}, \label{RealCol}
\end{equation}
where $N_J$ is proportional to the ``number of jets in the quark'', a DGLAP-type emission kernel. 

As shown in~\cite{Boussarie:2016ogo}, combining Equations~(\ref{VirtualDiv}), (\ref{RealSoft}), (\ref{RealCol}) and the equivalent of Equation~(\ref{RealCol}) where the gluon is collinear to the antiquark, one finally obtains a finite cross section.

\section{CONCLUSION}

Dijet production in DDIS at HERA was recently analyzed~\cite{Aaron:2011mp}, and a precise comparison of 
dijet versus triple-jet production, not performed yet at HERA~\cite{Adloff:2000qi}, would be of much interest. Investigations of the azimuthal distribution of dijets in diffractive photoproduction performed by ZEUS~\cite{Guzik:2014iba} show signs of a possible need for a 2-gluon exchange model, which is part of the shockwave mechanism. Our calculation could be used for phenomenological studies of those experimental results. Complementary studies could be performed at LHC with UPC events. Furthermore, our result could be extended to  diffractive exclusive meson photo- and electroproduction.

\section*{ACKNOWLEDGMENTS}

A.~V.~Grabovsky acknowledges support of president scholarship 171.2015.2, RFBR grant 16-02-00888, Dynasty foundation, Metchnikov grant and University Paris Sud. He is also grateful to LPT Orsay for hospitality  while part of the presented work was being done. R.~Boussarie thanks RFBR for financial support via grant 15-32-50219. This work was partially supported by the ANR PARTONS (ANR-12-MONU-0008-01), the COPIN-IN2P3 Agreement and the Th\'eorie-LHC France Initiative. R.~Boussarie and L.~Szymanowski were supported by grant No 2015/17/B/ST2/01838 of the National Science Center in Poland.

\end{document}